\documentclass[11pt]{article}

\usepackage{amsfonts,amssymb,color,cancel,mathtools,setspace,tabularx}
 \usepackage{sidecap}
  \usepackage{amsmath}
\usepackage{graphicx}
\usepackage{amssymb}
\usepackage{pifont}
\usepackage{ulem}
\usepackage{framed}
\usepackage{cite}
\newcommand{\ba}{\begin{eqnarray}}
\newcommand{\ea}{\end{eqnarray}}

\usepackage{hyperref}
\hypersetup{
 colorlinks=true, 
    linktoc=page,     
    linkcolor=blue,
}
\newcommand{\bea}{\begin{eqnarray}}
\newcommand{\eea}{\end{eqnarray}}
\newcommand{\be}{\begin{equation}}
\newcommand{\ee}{\end{equation}}

\newcommand{\st}[1]{\ifmmode\text{\sout{\ensuremath{#1}}}\else\sout{#1}\fi} 

\makeatletter

\begin{document}

\title{\LARGE First Law for Kerr Taub-NUT AdS Black Holes}

\author{
{\large Nelson Hernández Rodríguez}$^{2}$ \footnote{nelson.hernandez@estudiante.uam.es} and {\large Maria J. Rodriguez}$^{1,2}$ \footnote{maria.rodriguez@usu.edu, maria.rodriguez@csic.es}, \\
\\
$^{1}${\small  Department of Physics, Utah State University,}\\ {\small 4415 Old Main Hill Road, UT 84322, USA}\\
\\
\\ 
 $^{2}${\small  Instituto de F\' isica Te\' orica UAM/CSIC,}\\ { \small Universidad Aut\'  onoma de Madrid, Cantoblanco, 28049 Madrid, Spain} \\
\\
\\
 }

\maketitle

\abstract
The first law of black hole mechanics, which relates the change of energy to the change of entropy and other conserved charges, has been the main motivation for probing the thermodynamic properties of black holes. In this work, we investigate the thermodynamics of Kerr Taub-NUT AdS black holes. We present geometric Komar definitions for the black hole charges, that by construction satisfy the Smarr formula. Further, by a scaling argument based on Euler’s theorem, we establish the first law for the Kerr Taub-NUT AdS black holes. The corresponding first law includes variations in the cosmological constant, NUT charges and angular momenta. The key new ingredient in the construction are the independent variations of both angular momenta, the black hole and Misner string angular momenta. Employing the Brown-York quasi-local charge definitions we show that our expression for the mass and spin coincide with our generalized Komar expressions. We indicate the relevance of these results to the thermodynamics of rotating AdS black holes, including the proper choice of time-like Killing vector to produce the correct thermodynamic mass.

\newpage

\tableofcontents


\section{Introduction}

The identification of the entropy with $1/4$ of the event horizon area depends crucially on the fact that the first law of black hole mechanics holds. The relationship usually called the first law is actually the black-hole version of the fundamental identity of thermodynamics. Without this, we could merely give a hand-waving argument for the relationship among these quantities based on the idea that both the entropy and event horizon area are non-decreasing functions. This entropy/area relation has been found to be violated in the presence of a NUT charge \cite{Mann:1999bt,Astefanesei:2004kn}, and restored for some specific NUT black hole solutions via generalized Komar charge definitions and the first law \cite{Hennigar:2019ive, Bordo:2019slw,Kastor:2009wy}.

The properties of Taub-NUT black hole spacetimes and thermodynamics have been studied by many authors and in many different contexts \cite{Mann:1999bt,
Astefanesei:2004kn,Clarkson:2002uj,Clarkson:2003wa,Padmanabhan:2011ex,Johnson:2014xza,Clement:2015cxa,Araneda:2016iiy,Donnay:2019zif,Kol:2019nkc,Durka:2019ajz,Ciambelli:2020qny}. Recently, special attention has been paid to the presence of Misner strings or rotation \cite{Hennigar:2019ive, Bordo:2019slw,BallonBordo:2019vrn}. One of the long-standing problems in black hole physics involves describing the thermodynamic charges and first law of rotating Taub-NUT black holes in anti de Sitter (AdS) spacetimes. These spaces have non-trivial NUT charges which are associated to a Killing vectors that are not properly normalized at spatial infinity -- their norm is $\theta$-dependent. And yet, the situation is far subtler than in asymptotically flat spacetimes. For rotating black holes in AdS, there are also inherent issues associated with the proper choice of timelike Killing vector to produce the correct thermodynamic mass \cite{Gibbons:2004ai,Caldarelli:1999xj}. This led to a lack of unanimity about what precisely are the correct expressions for the total mass or energy for Kerr-AdS black holes to satisfy the first law \cite{Hawking:1998kw}. 

The fact that the angular velocity $\Omega_{\infty}$ does not vanish at infinity is a salient feature of rotating black holes in AdS space. Rather than the angular velocity on the black hole horizon $\Omega_{BH}$, it is easily verified that in the first law for AdS black holes one should use the angular velocity $\Omega=\Omega_{BH}-\Omega_{\infty}$, which is measured relative to a non-rotating observer at infinity. We will argue that this will also be the case for Kerr Taub-NUT AdS black holes.

Despite these successes for spacetimes with a negative cosmological constant, it is important to recognize that there are obvious difficulties in deriving the Smarr formula, and its variation. Various infinite integrals are encountered, and as we will find out here, even the evaluation of the Komar integrals and the proper definition of the generalized co-potentials requires care. Recently the first law was obtained for the asymptotically flat rotating NUT-black holes \cite{BallonBordo:2019vrn}. This was accomplished by introducing a choice of a potential and conjugate charge for the NUT parameter \cite{BallonBordo:2019vrn,Bordo:2019tyh}
and a generalization of the Komar-like integrals introduced in \cite{Bordo:2019tyh}. Notably, the angular momentum picks up non-trivial contributions from Misner strings and is no longer given by the Noether charge over the sphere at infinity. One then wonders if other approaches, including the Brown-York quasi local construction for the conserved charges can reproduce and extend the notions about the properties of the rotating NUT black holes in AdS.

Building upon previous analysis and results, one can then forsee that to render an intrinsic construction of the first law for Kerr Taub-NUT AdS the relation should be recast with additional contributions not only from the NUT charges, and Misner string angular momenta but also with proper expressions for the total mass and angular velocities.

In this paper we will carefully address these issues and identify the generalized first law rotating Taub–NUT AdS black hole solutions. We identify the temperature of the spacetime with the Hawking temperature of the black hole horizon, $T$, and the entropy of the spacetime with the entropy of the black hole horizon, $S$, and seek other thermodynamic charges so that the following generalized first law is satisfied: 
\be\label{eq:FL}
dM=T\,dS+\Omega_{BH}\, dJ-\Omega_{\infty} \,d J_{BH}+ \psi_+ \,dN_+ +\psi_-\, dN_- + V\, dP\;,
\ee
together with the corresponding Smarr relation
\be\label{eq:Smarr}
M=2 \, (T\,S+\Omega_{BH}\, J-\Omega_{\infty}\, J_{BH}+ \psi_+ \, N_+ +\psi_-\, N_- - P\, V)\;,
\ee
where M stands  for  the  mass  of  the  solution, $\Omega_{BH}$ and $\Omega_{\infty}$ are respectively the angular velocities of the black hole and the $AdS$ boundary, $J_{BH}$ the black hole angular momentum and $J$ the total black hole spin\footnote{The total black hole angular momenta $J$ will include contributions from the bulk black hole and Misner string as in \cite{BallonBordo:2019vrn}.}, and $N_{\pm},\psi _{\pm}$ are respectively the  gravitational  Misner charges and independent conjugate potentials. 
Obviously, such a first law is of full cohomogeneity: it has 6 terms which correspond to the variation of the 4 physical parameters of the solution: $\{r_+, a, n, l\}$. 
This relation also involves P, the cosmological constant pressure and V is the conjugate thermodynamic volume,

\be
P=-\frac{\Lambda}{8\pi},\qquad V=\left(\frac{\partial M}{\partial P}\right)_{S, N}\,.
\ee
In the above discussion, we treated the cosmological constant as a thermodynamic state variable through the pressure $P$. Had we treated the cosmological constant as a fixed parameter, $ V d P$ term would vanish, and yet the first law of thermodynamic(\ref{eq:FL}) would hold.

Our approach, in contrast to other studies will also include the Brown-York quasi-local definitions for the thermodynamic charges. The construction then defines the generalized first law of thermodynamics for the rotating Taub-NUT AdS black hole solutions with additional terms that organize the information about  the  solutions in fascinating new ways.

To study the phase structure and stability of black holes one can analyze the potentials and response functions in different thermodynamic ensembles. Another application of our work is in understanding the grandcanonical ensemble. Semiclassically, the total action is evaluated from the classical solution of field equations, yielding an expression for the Gibbs-Duhem relation  to the partition function \cite{GibbonsHawking}. The Euclidean action $I$ is calculated with the usual counter terms see e.g. \cite{Emparan:1999pm}. It is associated with a Gibbs free energy, ${\cal G}=I/\beta$, where $\beta$ is the periodicity of the Euclidean time coordinate. The relation 
\bea\label{eq:Gibbs}
{\cal G}(T, \Psi_{\pm},P,\Omega_{BH},\Omega_{\infty})=M- T\,S-\Omega_{BH}\, J+\Omega_{\infty}\, J_{BH}- \psi_+ \, N_+ -\psi_-\, N_- \,,
\eea
is equivalent to the Smarr formula. Previously there has been one other proposals for thermodynamic potentials for Kerr Taub-NUT AdS put forward by \cite{Sharif:2021zmr} which differs from our expressions. The discrepancy is due to the different choice of normalizations of the Killing fields.

The remainder of this article is organized as follows:\\
In section \ref{sec:RotAdSMetric} we review the four-dimensional Kerr Taub-NUT AdS black hole, give the definitions of some of the thermodynamics quantities like entropy and temperature, and recall the conditions under which some symmetries are preserved. In section \ref{sec:charges} a generalized Smarr relation, by identifying the Komar charges of the Kerr NUT AdS black hole. We also compute the corresponding Brown-York charges for the rotating Taub-NUT AdS black holes, and extend the results of the Brown-York thermodynamic charges for the Kerr Taub-NUT AdS black hole with a non-symmetrical distribution of the Misner strings. The first law of thermodynamics is considered in section \ref{sec:first law}. Finally, section \ref{sec:discussion} contains our conclusions. Throughout the paper, we use geometric units with $G=\hbar=c=1$ and the metric signature $(-,+,+,+)$.

\section{Rotating AdS Taub-NUT black hole}
\label{sec:RotAdSMetric}

We begin by reviewing the rotating AdS Kerr Taub-NUT black hole solution. The metric in Boyer-Lindquist type coordinates \cite{Griffiths:2005qp} is given by
\ba\label{NUT2}
ds^2&=&-\frac{\Delta}{\Sigma}\Bigl[dt+(2n\cos \theta-a\sin^2\!\theta)\frac{d\phi}{\Xi}\Bigr]^2+\frac{\Delta_{\theta}}{\Sigma}\Bigl[a\,dt-(r^2+a^2+n^2)\frac{d\phi}{\Xi}\Bigr]^2\nonumber\\
&&+\frac{\Sigma}{\Delta}\,dr^2+\frac{\Sigma}{\Delta_{\theta}}\sin^2\!\theta\, d\theta^2\,,
\ea
where
\ba
\Sigma&=&r^2+(n+a\cos\theta)^2\,,\\
\frac{\Delta_{\theta}}{\sin^2\!\theta}&=&\,1-\frac{4an\cos\theta}{l^2} -\frac{a^2\cos^2\!\theta}{l^2}\,,\label{co}\\
\Delta&=&\!r^2+ a^2\!-\!2mr\!-\!n^2\!+\frac{3(a^2\!-\!n^2)\,n^2+({a^2}\!+\!6n^2)\,r^2\!+\!{r^4}}{l^2}\;.
\ea
and $\Xi=1-(a^2/l^2)$. The space-time (\ref{NUT2}) is a solution of Einstein's equations $R_{\mu\nu}-(1/2)\,R\, g_{\mu\nu}+\Lambda \, g_{\mu\nu}=0$ with negative cosmological constant $\Lambda=-3/l^2$, and $l$ the AdS length. The mass and angular momentum are related to the parameters m and a appearing in
the metric, and $n$ is related to the NUT charge.

The spacetime is stationary and axisymmetric, corresponding to the Killing vectors
\ba
\xi_t=\partial_t/\Xi, \qquad \xi_{\phi}=\partial_{\phi}
\ea
which will be associated to conserved charges. Stirring at the expression of the  Killing vectors one can recognize a new function $\Xi$ in $\xi_t$ that fixes the normalization of the Killing vector as in Ref. \cite{Kostelecky:1995ei} so that the corresponding conserved quantities generate the algebra $so(3, 2)$. We will return to the expressions of the conserved charges and emphasize the importance of the normalization in section \ref{sec:GSmarr}.

This space-time also admits a number of Killing horizons. A study of the zeroes of $\Delta$ show that the for $a<l$ the metric represents a black hole, with an event horizon at $r=r_+$, where $r_+$ is defined as the largest root $\Delta(r_+)=0$. The black hole horizon is generated by a (properly normalized at infinity) Killing vector
\ba\label{eq:xi}
\xi=\partial_t+\Omega_{BH}\, \partial_{\phi},
\ea
where $\Omega_{BH}$ is the angular velocity evaluated at $r_+$.

Finally, as discussed in  \cite{Bordo:2019tyh, Misner} there are other Killing vectors for  the spacetime.  Namely,  when  the  Misner  string  is  present, the north/south pole axis is a Killing horizon of the following Killing vector
\ba
\xi_{\pm}=\partial_t\mp \frac{\Xi}{2 n}\partial_{\phi}.
\ea
Contrary to all other Killing  vectors $\xi_{\pm}$ are not properly normalized at infinity and their norm is there $\theta$-dependent.

\section{Charges and generalized Smarr relation}
\label{sec:charges}

In order to define a generalized Smarr relation and first law for rotating Taub-NUT AdS black holes, we must first discuss the conserved charges which are involved.  In this section we compute these relations and charges with two different methods: Komar integrals and the Brown-York formalism, also known as the quasi-local formalism.

\subsection{Thermodynamic quantities}

The basic assumption in our derivation of the first law of thermodynamics is to identify the entropy with the black hole horizon area $A$ according to the Bekenstein–Hawking entropy
\bea
S&\equiv& \frac{A}{4}= \int_0^{2\pi} \int_0^{\pi} \sqrt{g_{\theta\theta} \, g_{\phi\phi}} \, \bigg|_{r_+} d\theta d\phi
=\,\pi\, \frac{(r_+^2+a^2+n^2)}{\Xi}\,,
\eea
and, by standard arguments, the Hawking temperature of the black hole horizon through the surface gravity $\kappa$ such that
\bea
T&\equiv&\frac{\kappa}{4\pi}=\frac{\Delta'(r_+)}{4\pi(r_+^2+a^2+n^2)}\,.
\eea

Another quantity of interest in the first law of thermodynamics is the angular velocity $\Omega$. The fact that the angular velocity does not vanish at infinity $r\rightarrow \infty$ for the solution (\ref{NUT2}) with $m=n=0$ is a salient feature of rotating black holes in AdS space, different from the asymptotically flat case, where $\Omega_{\infty}=0$. This definition became key in deciphering some years ago a first law for rotating AdS black holes (without NUT charge) \cite{Gibbons:2004ai,Caldarelli:1999xj,Hawking:1998kw}. The angular velocity entering the thermodynamics is measured relative to a non-rotating frame at infinity, and is given by the difference
\bea
\Omega&=&\Omega_{BH}-\Omega_{\infty} = \frac{a \, (1+(r_++n^2)\,l^{-2})}{r_+^2+a^2+N^2}\,,
\eea
where $\Omega_{BH}$ represent the angular velocity of the event horizon
\bea
\Omega_{BH}=-\frac{g_{t\phi}}{g_{\phi\phi}}\bigg|_{r_+}=\frac{a \, \Xi}{r_+^2+a^2+n^2}\,,
\eea
and
\bea
\Omega_{\infty}=-a/l^2\,,
\eea
the angular velocity of the boundary AdS space-time.
The fact that the angular velocity relevant for the rotating AdS NUT black hole turns out to be the one of the Einstein universe in AdS boundary, agrees nicely with the AdS/CFT correspondence.

In addition to the black hole, the metric (\ref{NUT2}) contains Misner strings attached to the poles of the black hole (at $\theta=0$ and $\theta=\pi$) that extend to infinity. One can therefore find non trivial surface gravities due to these Misner strings. These can be calculated using the standard formula $\kappa_{\pm}^2=\nabla_{\mu}L\nabla^{\mu}L/ (4L)$, $L=-\xi_{\pm}^2$. In what follows we shall associate with these quantities the following ‘Misner potentials’
\ba
\psi_{\pm}=\frac{\kappa_{\pm}}{4\pi}=\frac{\Xi}{8\pi n}.
\ea

%
Finally, as shown in previous works e.g. \cite{Caldarelli:1999xj, Bordo:2019tyh}, the first law for rotating AdS black holes will include a pressure
\bea
P=-\frac{\Lambda}{8\pi}\,
\eea
that promotes the cosmological constant to a thermodynamic state variable.

The remaining thermodynamic charges -- including the physical mass, angular momenta, Misner charges and volume -- are computed in the following sections employing the generalized Komar integrals and Brown-York charges.

\subsection{Generalized Komar integrals and Smarr relation}
\label{sec:GSmarr}

To find the Komar integral prescription for the Misner charges, and the angular momentum $J$ we derive a generalized Smarr relation for the Kerr-Taub-NUT AdS space-time and identify the due Komar integrals with the thermodynamic charges.

We follow  \cite{Bordo:2019tyh} and the prescription to include rotation. Our starting point is the black hole horizon $\xi$ defined in (\ref{eq:xi}). The relation for any Killing vector implies that at least locally there exists a Killing co-potential 2-form $\omega$:
\bea\label{eq:killing1}
\nabla_{\mu} \xi^{\mu}=0 \Rightarrow \xi^{\nu}=\nabla_\mu \omega^{\mu\nu}
\eea
which is set by the Conformal Killing-Yano (CKY). The co-potential, however is not uniquely defined, and one can always perform a shift. This will become relevant for our computations later.
Yet, this other relation for the Killing vector is satisfied
\bea
\nabla_{\nu}\nabla^{\nu} \xi_{\mu}+R_{\mu\nu}\xi^{\nu} =0 
\eea
that for an Einstein space $R_{\mu\nu}-\Lambda g_{\mu\nu}=0$ combined with $(\ref{eq:killing1})$ yields
\bea
\nabla_{\nu}\nabla^{\nu} \xi_{\mu}+\Lambda \nabla_{\nu}\omega^{\nu}_{\mu} =0 
\eea
or equivalently in differential forms
\bea\label{integral}
0=\int_{\Sigma} d*\left(d\zeta+2\Lambda \omega \right)=\int_{\partial\Sigma}  *\left(d\zeta+2\Lambda \omega \right),
\eea
integrating over the $3$-dimensional (t = constant) hypersurface $\Sigma$, and $\partial \Sigma$ the corresponding spacetime boundary.
This expression relies on the existence of a Killing co-potential 2-form $\omega$ such that for the generator of the black hole horizon $\zeta=k+\Omega_{BH}\,\eta$ verifies $k^a=\nabla_b\omega^{ba}$. See  \cite{BallonBordo:2019vrn}  for more details. Here, to simplify the notation we label the Killing vectors $k=\partial_t$ and $\eta=\partial_{\phi}$. 
Before computing the different terms in (\ref{integral}), note that the Killing potential is not unique. If a co-potential $\bar\omega$ solves $k^a=\nabla_b \bar \omega^{ba}$, so will ${\omega}^{ab}=\bar\omega^{ab}+\lambda^{ab}$, as long as $\nabla_a \lambda^{ab}=0$. We will show in the following section, that the co-potential in \cite{Kubiznak:2007kh} for Kerr Taub-NUT AdS spacetime will require a proper gauge choice $\lambda$ to produce the correct thermodynamic mass.

To find the generalized Smarr law, we first define the boundary $\partial\Sigma$. The boundary $\partial\Sigma$  thus consists of the two Misner string tubes $T_+$ and $T_-$ located at $\theta=\epsilon$ and $\theta=\pi-\epsilon$ respectively with $\epsilon \rightarrow 0$, the black hole horizon H at $r=r_+$, and the
sphere $S_{\infty}$ at $r=\infty$. Taking into account the orientation of the boundaries we have
\bea
\partial\Sigma =T_+ + S_{\infty}-T_- + H\,.
\eea

The integral (\ref{integral}) thus splits into the following six contributions:
\bea\label{GenSmarr}
0&=&\int_{S_{\infty}} *\left( dk+ 2 \Lambda \omega \right)- \int_H *d\zeta+\Omega_{BH} \left(\int_{S_{\infty}}  *d\eta+\int_{T_+} *d\eta-\int_{T_-} *d\eta\right)\\
&&+\left(\int_{T_{+}}*dk+ 2 \Lambda \Omega^{(r=\infty)}_{+}\right)-\left(\int_{T_{-}}*dk+ 2 \Lambda \Omega^{(r=\infty)}_{-}\right)\\
&&-2\Lambda\left(\int_H *\omega+\Omega_+^{(r=r_+)}-\Omega_-^{(r=r_+)}\right).
\eea

Here, the integration over the Misner tubes $\int_{T_{\pm}}$ is over $r \in [r_+,\infty]$, $\phi\in[0,2\pi]$ and at fixed $t = const$. We have also introduced the notation
\bea
\int_{T_{\pm}} *\omega=\Omega^{(r=\infty)}_{\pm}-\Omega^{(r=r_+)}_{\pm}\,.
\eea

The first term in (\ref{GenSmarr}) yields the thermodynamic mass $M$ and includes an additional term involving the angular velocity $\Omega_{\infty}$ 
\bea\label{mass}
M+2 \, \Omega_{\infty} J_{BH}&=&-\frac{1}{8\pi}\int_{S_{\infty}} *\left( dk+ 2 \Lambda \omega \right),
\eea
where $J_{BH}=a\,M/\Xi^2$ is the black hole angular momentum corresponding to the $\eta$ Killing vector. The term involving $\Omega_{\infty}$ is new, and vanishes in the limit of zero rotation $a\rightarrow 0$ as in \cite{Bordo:2019tyh}. 

The arrangement of the other terms is more standard. A contribution comes from the second term that gives the product of entropy and temperature
\bea
TS&=&-\frac{1}{16\pi} \int_H *d\zeta.
\eea
The third term define the total angular momentum
\bea
J=\frac{1}{16\pi}\left(\int_{S_{\infty}}  *d\eta+\int_{T_+} *d\eta-\int_{T_-} *d\eta\right)\,,
\eea
that includes contributions from the black hole's angular momenta $J_{BH}$ and Misner's strings $J_s$.\\ 
The next two terms define the gravitational Misner charges:
\bea\label{MisnerCharges}
\psi_{\pm} N_{\pm}=\pm\frac{1}{16\pi}\left(\int_{T_{\pm}}*dk+ 2 \Lambda \Omega^{(r=\infty)}_{\pm}\right)\,.
\eea
The final integral yields the modified thermodynamic volume
\bea\label{volume}
V=-\left(\int_H *\omega+\Omega_+^{(r=r_+)}-\Omega_-^{(r=r_+)}\right).
\eea
One can further readily verify, replacing all the above definitions (\ref{mass})-(\ref{volume}) into (\ref{GenSmarr}), that by construction, the defined quantities obey the generalized Smarr relation
\bea\label{eq:Smarr}
M=2 \, (TS+\Omega_H J-\Omega_{\infty} J_{BH}+ \psi_+N_+ +\psi_-N_- - P V)\;.
\eea
It may at first seem strange to consider additional $\Omega_{\infty} J_{BH}$  contributions in the definition of the mass. Let us note, however, that appearance of the angular velocity at infinity itself in the Smarr relations has precedence in the literature \cite{Gibbons:2004ai,Caldarelli:1999xj}. Moreover, that the angular velocity relevant to AdS black hole thermodynamics turns out to be the one of the rotating Einstein universe at the AdS boundary, agrees nicely with the AdS/CFT correspondence; if the AdS black hole in the bulk is described by a conformal field theory living on the boundary, then the relevant angular velocity entering the thermodynamics should be that of the rotating Einstein universe at infinity. The prescription for mass M and other quantities is non-unique. The ultimate check is the validity of the first law of thermodynamic. This will be the focus in Section \ref{sec:first law}. To justify these prescription the best one can do is to choose such mass $M$ so that the integral (\ref{mass}) yields the true physical mass  (\ref{eq:Smarr}). 

In the following, we are going to use these generalized Komar-like definitions to compute the charges associated with the Kerr Taub-NUT black hole metric (\ref{NUT2}).

\subsubsection*{Generalized Smarr relation for AdS-Kerr-NUT}

Before diving into the explicit computations of the Komar integrals, we discuss the two form dual to the co-potential, $ *\omega $ which is involved in these definitions. The Kerr Taub-NUT AdS metric admits a hidden symmetry encoded in two tensors: the principal Killing–Yano (KY) tensor $\textbf k$ and dual Conformal Killing-Yano (CKY) tensor $\textbf h= * {\textbf k}=d \textbf b$ as shown in \cite{Kubiznak:2007kh}. Both of these tensors have wide applications in physics related to hidden (super)symmetries, conserved quantities, symmetry operators, or separation of variables. In our computations, the KY tensor plays a central role in the definitions of the generalized Komar charges as we have shown in the preceding section. The CKY tensor sets the Killing co-potential $\omega=\textbf h/3$ for the Killing vector $k$. For the  Komar integrals one can directly compute $* \omega= *\textbf h/3= **\textbf h/3  = - \textbf k /3$ such that \footnote{One needs to shift the coordinates in \cite{Kubiznak:2007kh} by $\phi^{there}\rightarrow\phi/\Xi$  and $\tau^{there}\rightarrow t+ 2 n \phi/\Xi$ to match with the coordinate system employed in our paper.}
 \bea\label{omega}
 *\omega &=&-\frac{1}{3}(n+a \cos\theta) dr \wedge\left[dt+\frac{d\phi}{\Xi}(2 n \cos\theta- a \sin^2\theta)\right]  \\
&& +\frac{1}{3} r \sin\theta d\theta \wedge \left[ a dt-\frac{d\phi}{\Xi} \left(n^2+a^2+r^2+ \frac{ m a^2}{ r \, \Xi} \right)\right]\,.\nonumber
 \eea
Of course the co-potential $\bar\omega$ defined by means of Killing-Yano tensor is not unique; one may add, as we briefly outlined earlier, any co-closed 2-form $\lambda$ to $\bar\omega$. The construction in \cite{Cvetic:2010jb} focused on building the correct mass for the Kerr-AdS black holes (carrying no NUT charge ($n=0$)) \cite{Gibbons:2004ai,Caldarelli:1999xj}. The authors found that one has to add to the co-potential in \cite{Kubiznak:2007kh} a gauge correction $\lambda$. In turn, when the black hole contains a non trivial NUT charge, the corresponding gauge correction translates in a contribution to $*\bar\omega$ such that
\bea
 *\omega \rightarrow  *\bar\omega +*\lambda
 \eea
with $*\lambda= -\frac{1}{3} \sin\theta \frac{ m a^2}{ \, \Xi^2} d\theta \wedge {d\phi} $ found in (\ref{omega}).
  
Now, using the above expressions of the two form components, we can evaluate the generalized Komar integrals (\ref{mass})-(\ref{volume}). We find the following mass and angular momenta:
\bea
M&=&\frac{m}{\Xi^2}\,,\qquad J_{BH}=\frac{a\, m}{\Xi^2} \\
J&=&\frac{a\,(a^2+n^2+r_+^2)(1+(3n^2+r_+^2)/l^2)}{2 r_+\, \Xi^2 }\,.
\eea
Here $J$ is the total angular momentum defined as the sum of the Misner string $J_s$  and the black hole angular momentum $J_{BH}$, such that $J=J_s+J_{BH}$. In particular, the total angular momenta $J$ differs from the asymptotic charge $J_{BH}$ by the Misner string contribution
\bea
J_s=\frac{a\,n^2 (1+(3n^2-r_+^2)/l^2)}{ r_+\, \Xi^2 }\,.
\eea
 
Considering the integrals over the Misner strings
\bea
\int_{T_{\pm}} *dk&=&-\lim_{\epsilon \rightarrow 0} \int_{r=r_{+}}^{\infty}(*dk)_{r\phi}\, dr\,d\phi\\
&=&-8 \pi n \left[\frac{ r^3 (a\pm n)+r \left(a^3\pm 4 a^2 n+3 a n^2\mp n \left(l^2+3 n^2\right)\right)+l^2 m (a\pm n)}{l^2\,\Xi\, \left(r^2+(a\pm n)^2\right)}\right]^{r\rightarrow\infty}_{r=r_+} \\
\Omega^{(r=\infty)}_{\pm}&=& - \lim_{r\rightarrow\infty}\frac{4\pi n}{3\,\Xi}\,  (a\pm n)\, r\,,
\eea
the Misner charges (\ref{MisnerCharges}) yield
\bea
\psi_{\pm}N_{\pm}&=&-\frac{n}{4r_+\Xi}\left[(n\mp a)\left(1+\frac{3n^2}{l^2}\right)-3(n\pm a)\frac{r_+^2}{l^2}\right]
\eea
and the volume:
\bea
V&=&\frac{4\pi}{3\,\Xi} \left( r_+\, (r_+^2+3n^2+a^2) + \frac{  a^2 m}{\Xi} \right)\,.
\eea
For a vanishing NUT parameter, $n=0$, the volume $V\rightarrow \Theta^{there}(-8 \pi)$ becomes precisely the corresponding conjugate variable to the cosmological constant $\Theta^{there}$ defined in \cite{Caldarelli:1999xj} for Kerr-AdS black holes.

All these quantities  that we computed here add up precisely to fulfill the relation (\ref{eq:Smarr}). With these geometric quantities in hand, let us now turn towards the definitions of the Brown-York charges.

\subsection{Brown-York charges}
\label{secBY}

In the previous section we have presented the definitions for the physical conserved quantities by evaluating Komar integrals. Instead, one can compute the conserved charges by the direct use of the finite action, applying the method developed by Brown and York \cite{BrownYorky}.

Let us briefly review how this formalism works. The basic idea of Brown and York was to define quasilocal charges by enclosing a given region of spacetime with some surface on which one can perform the charge computations. Provided one incorporates appropriate boundary terms it is possible to extend the quasilocal surface to spatial infinity without difficulty. From a given finite action $I$ supported with counterterms, one  can derive the local surface energy-momentum stress tensor 
\bea\label{stresstensor}
 \tau_{ab} =\frac{2}{\sqrt{-h}}\frac{\delta I}{\delta h^{ab}}
\eea
 with induced metric $h_{ab}$. The induced metric can be written, at least locally, in ADM-like form
\bea
h_{ab} dx^a dx^b= - N^2 dt^2+\sigma_{ab}(dy^a+N^a\, dt)(dy^b+N^b\, dt)\,,
\eea
where $N$ and $N^a$ are the lapse function and the shift vector respectively. We indicate by $u^a$ the unit normal vector of a spacelike hyper-surface $S_t$ at constant t and by $\partial  \mathcal{M}$ the spatial boundary of the spacetime manifold $\mathcal{M}$. Moreover, $\Sigma=S^2$ is the spacelike intersection $S_t \cap \partial  \mathcal{M}$ embedded in $\partial  \mathcal{M}$ with induced metric $\sigma_{ab}$.
Then, for any Killing vector field $\xi_a$ associated with an isometry of the boundary three-metric, one defines the conserved charge
\bea\label{charge}
Q_{\zeta}=\int_{\Sigma} d^2x \sqrt{\sigma} u^a \tau_{ab} \xi^b\,.
\eea
We turn now to the calculation of the conserved Brown-York charges. In the case of the Kerr-Taub NUT AdS geometry, the Euclidean action is calculated with the usual counter terms \cite{Emparan:1999pm}:
\bea\label{action}
I&=&I_{bulk}+I_{surface}+I_{ct}\,,
\eea
such that
\ba
I_{bulk}&=&\frac{1}{16\pi}\int_{ \mathcal{M}}d^{4}x\sqrt{g}\left( R+\frac{6}{ l^{2}}\right)\,\\
I_{surface}&=& \frac{1}{8\pi}\int_{\partial  \mathcal{M}}d^{3}x\sqrt{h}\left[\mathcal{K}\right]\,\\
I_{ct}&=&- \frac{1}{8\pi}\int_{\partial  \mathcal{M}}d^{3}x\sqrt{h}\left[  \frac{2}{ l} + \frac{ l}{2}\mathcal{R}\left( h\right) \right]\,.
\ea
The first term in eq. (\ref{action}) is the bulk action over the 4-dimensional manifold $ \mathcal{M}$ with metric $g$ and the second term is the surface term necessary to ensure that the Euler-Lagrange variation is well-defined. Here, $h$ is the induced metric of the boundary,  $\mathcal{K}$ is the extrinsic curvature and $\mathcal{R}\left( h\right)$ the Ricci scalar of the boundary. 

We follow the definitions for the Brown-York tensor (\ref{stresstensor}) in \cite{Caldarelli:1999xj}. The application of the action (\ref{action}) gives a divergence-free boundary stress tensor
\bea
 \tau_{ab} =\frac{2}{\sqrt{-h}}\frac{\delta I}{\delta h^{ab}}= -\frac{1}{8\pi G}\left[(\mathcal{K}_{ab}-h_{ab}\, \mathcal{K}) + \frac{2}{l}h_{ab} -l (\mathcal{R}_{ab} -\frac{1}{2} h_{ab}\, \mathcal R)\right]\,,
\eea
where $\mathcal{R}_{ab}$ is the Ricci curvature  tensor of the (arbitrary) boundary $ \partial \mathcal{M}$.
 In the case of Kerr Taub-NUT AdS black hole solution, we choose the $\partial \mathcal{M}$ to be a three-surface of fixed $r$. In turn, we obtain the normal vectors
 \bea
 n_{\mu}=\sqrt{g_{rr}} \, \delta^r_{\mu} ,\qquad  u_{\mu}=- N\, \delta^t_{\mu}\,,
 \eea
 where $N$ is the lapse function defined by $N=\sqrt{\frac{\rho^2 \Delta \Delta_{\theta}}{(r^2+a^2)^2\Delta_{\theta}-a^2 \Delta \sin^2\theta}}$ satisfying the normalization condition $u . u= -1$.
 

 At the boundary $r=\infty$ we obtain
\bea
 8\pi G \tau_{tt} &=&\frac{2 m }{r l } +\mathcal{O}\left(\frac{1}{r^2}\right)\,.\\
  8\pi G \tau_{t \phi} &=&-\frac{2 m (a \sin^2 \theta - 2 n \cos\theta)}{r l \Xi}  +\mathcal{O}\left(\frac{1}{r^2}\right)\,,\\
   8\pi G \tau_{\phi \phi} &=&\frac{ m }{r l \Xi^2} \left[\sin ^2 \theta \left( l^2+a^2   (3 \sin ^2\theta- 1 )-12 a  n \cos \theta \right)+8 n^2 \cos^2 \theta\right]+\mathcal{O}\left(\frac{1}{r^2}\right)\,,\\
    8\pi G \tau_{\theta \theta} &=&\frac{m l }{r \Delta_{\theta} }+\mathcal{O}\left(\frac{1}{r^2}\right)\,,
\eea
all the other components vanish. Combining the last equations with (\ref{charge}) gives the quasi-local energy and angular momenta respectively
\bea \label{eq:BYcharges}
M\equiv Q_{\partial_t/\Xi}=\frac{m}{\Xi^2} \,,\qquad J_{BH}\equiv Q_{\partial_{\phi}}=\frac{a m}{\Xi^2}\,.
\eea

Within the Brown-York approach that we are considering, we can also compute the Euclidean action yielding the thermodynamic potential relevant to the grand-canonical ensemble. More precisely, the action can be associated with the Gibbs free energy, ${\cal G}=I/\beta$, where $\beta$ is the periodicity of the Euclidean time coordinate. 
Evaluating the action (\ref{action}) on shell, one can easily find that there is a nontrivial bulk contribution - first two terms in (\ref{action}) - that simplifies to
\bea
I_{bulk}^{on shell}=\frac{1}{16\pi}\int_{ \mathcal{M}}d^{4}x\sqrt{g}\left(\frac{6}{ l^{2}}\right).
\eea
In the computation of the action one usually encounters infrared divergences as we have argued, the Brown York technique consists in adding suitable counterterms to the action that will cancel such divergences. These counterterms are built up with curvature invariants of a boundary $\partial  \mathcal{M}$ (which is sent to infinity after the integration), and thus obviously do not alter the bulk equations of motion. One then chooses the boundary $\partial  \mathcal{M}= S^1 \times S^2$, $S^1$ being the time circle and $S^2$ being a 2-sphere with a large radius, which has to be sent to infinity after the integration. The metric is stationary, hence the time integration gives rise to a simple multiplicative factor $\beta$. The integration requires a little bit of work, but anyway it can be performed in a closed manner.  This gives
\be\label{G}
{\cal G}=\frac{m}{2\,\Xi}-\frac{r_+(a^2+r_+^2+3n^2)}{2\,\Xi \,l^2}\,,
\ee
in complete agreement with \cite{Bordo:2019tyh}, and also \cite{Gibbons:2004ai,Caldarelli:1999xj} for $n=0$. Our computation can be considered as an independent check of the thermodynamic charges and potentials, the quantum statistical relation as well as the first law of thermodynamics. To cross check our results, let us turn towards the Euclidean action and the corresponding derivation of the thermodynamic quantities. By computing the Gibbs-Duhem free energy via (\ref{eq:Gibbs}), we find that it is consistent with the expected expression (\ref{G}).

\subsection*{General Distribution of Misner strings}
The Kerr Taub-NUT AdS metric (\ref{NUT2}) can also exhibit yet another dimensionful physical parameter $s$, which governs the overall distribution and strength of the Misner string singularities.  The parameter $s$ can be formally added by performing the ‘large coordinate transformation’
\bea
t\rightarrow t+ (2 s / \Xi) \, \phi\,.
\eea
In particular, when $s = +n$, there is only one Misner string located on the north(south) pole $\cos \theta = +1$($\cos \theta = -1$) axis, while the south(north) pole $\cos \theta = -1$($\cos \theta = +1$) axis is completely regular. For $s = 0$, we recover (\ref{NUT2}), where the two Misner strings are ‘symmetrically distributed’ and both axes are ‘equally singular’. We have extended the Brown-York quasi-local charge formulas (\ref{eq:BYcharges}) for a symmetric distribution of Misner strings (the case of $s = 0$ studied in Section \ref{secBY}) to a more general formula allowing for the asymmetric distribution of Misner strings with $s\ne 0$. The mass and angular momenta are then
\bea
M=\frac{m}{\Xi^2}\,,\qquad J_{BH}=\frac{m}{\Xi^2} (a- 3 s)\,.
\eea
The parameter s governing the Misner string distribution can be now freely varied in the first law. Properties of the corresponding first law, which allows for variable Misner string strengths, will have higher cohomogeneity than (\ref{eq:FL}). We leave these issues for future investigations.

\section{First Law}
\label{sec:first law}

In this section we consider the first law of black hole mechanics including a cosmological constant for the rotating black hole with NUT charges. Given the Komar construction for the thermodynamic quantities discussed in the previous section it is not difficult to verify that these quantities satisfy the desired first law
\bea
dM=T\,dS+\Omega_{BH}\, dJ-\Omega_{\infty} \,d J_{BH}+ \psi_+ \,dN_+ +\psi_-\, dN_- + V\, dP\;,
\eea
where all 4 parameters of the solution: ${n, r_+, a, l}$ may now be independently varied. Thus, we have shown (for the first time ever) how to formulate consistent thermodynamics of rotating Taub-NUT AdS solutions. 


In our discussion thus far, we have not made use of the first law of black hole mechanics for rotating Taub-NUT AdS black holes. It is widely believe that the first law is constructed for an observer whose asymptotic motion aligns with the generator of the black hole horizon, and not the generators of the string horizons. Yet, the relation (\ref{eq:FL}) does involve the angular velocity relative to a rotating observer at infinity $\Omega_{BH}$ for the conjugate black hole and Misner string angular momenta, and the boundary angular velocity $\Omega_{\infty}$. In the coordinates we are using, the angular velocity at the event horizon $\Omega_{BH}$ is (\ref{omega}). However, in asymptotically AdS spacetimes the conjugate variable to the angular momentum that enters the thermodynamics is the angular velocity of the horizon relative to the boundary $\Omega=\Omega_{BH}-\Omega_{\infty}$ where $\Omega_{\infty}=a/l^2$. Rearranging the terms in the first law, one can find its explicit expression of $\Omega$ relative to a frame that is non-rotating at infinity 
\bea
dM=T\,dS+\Omega_{BH}\, dJ_s+\Omega \,d J_{BH}+ \psi_+ \,dN_+ +\psi_-\, dN_- + V\, dP\;.
\eea
This means, that the Misner string rotates rigidly with the black hole event horizon. In contrast, the black hole rotates with a different angular velocity $\Omega$ as seen from an asymptotic observer. In the case without NUT charge, for $n=0$, the first law reduces to the results for AdS-Kerr in \cite{Gibbons:2004ai,Caldarelli:1999xj}.

\section{Discussion}
\label{sec:discussion}

In this paper, we have investigated some of the consequences of the rotation for Taub-NUT black holes with a cosmological constant, and showed how to formulate a consistent first law of rotating Taub– NUT AdS black hole solutions. Since the cosmological constant can be thought of as a pressure, this means that its conjugate variable in the first law is proportional to a volume. Yet, the rotation should also be treated as a thermodynamic variable in the first law of thermodynamics for black holes. In particular this involved different choices of normalizations of the Killing fields. The thermodynamics of the Taub–NUT spacetimes with the Misner strings present can be consistently formulated by adding a new pair of conjugate quantities $\psi_{\pm} N_{\pm}$. These additional assumptions enable the appropriate extension of the first law of black hole mechanics.

Using generalized Komar integrals, we have found proper definitions for the Kerr Taub-NUT AdS black holes. This procedure, is a generalization of the Komar method for asymptotically flat black holes, and involves the introduction of a Killing potential 2-form $\omega$. There is also a gauge freedom to add a closed 2-form to $\omega$. The key ingredient for the generilized charge definitions is to introduce a proper gauge choice to reproduce the physical mass compatible with a Smarr relation and first law. In addition, we have introduced the proper choices of normalizations of the Killing fields giving rise to the angular velocity definitions for the black hole and Misner string charge. Note that all  the thermodynamic quantities combined with their corresponding potentials have a smooth zero NUT charge limit, $n\rightarrow 0$, yielding thus a consistent thermodynamics of Kerr AdS spacetime. Having calculated all the thermodynamic quantities for these black hole solutions we have derived the Smarr relation and first law for the rotating Taub-NUT AdS black hole. 

Although we focused on the Komar definitions for the charges, generalizations involving Brown-York quasi-local charges are possible. We have presented the method for constructing the Brown-York quasi-local charges for the Kerr Taub-NUT AdS black hole solutions. The expressions that we have computed for the Brown-York mass and spin coincide with our generalized Komar expressions. This method allows for the precise identification of the black hole angular momenta $J_{BH}$. However, the method does not capture the Misner string angular momenta $J_s$. This suggests that new Brown-York charges associated with the Killing vectors $\xi_{\pm}$ may be defined to describe $J_s$ for the Misner strings. 

Let us conclude here with a short discussion about the {\it Reverse Isoperimetric Inequality} -- a geometric inequality involving the perimeter of a set and its volume. For black holes with or without rotation and/or charge, there are strong indications that the event horizon area $A$ (or 4 times the entropy S, such that $A=4S$ for the set) and the thermodynamic volume V, respectively, violated the isoperimetric inequality  \cite{Cvetic:2010jb}. This led to conjecture that all 4-dimensional black holes satisfy the {\it Reverse Isoperimetric Inequality}
\bea\label{eq:Ineq}
\left(\frac{3\,V}{\mathcal{A}_2}\right)^{1/3} \left(\frac{\mathcal{A}_2}{A}\right)^{1/2} \geq 1\,,
\eea
where $\mathcal{A}_2= 2\pi$ is the area of the unit sphere. It is interesting to examine whether or not the Kerr Taub-NUT AdS black holes obey this inequality. By a straightforward numerical inspection of (\ref{eq:Ineq}), our results imply that in fact they do obey the inequality, then the situation is somewhat analogous to the no-NUT charge AdS black holes.

Lastly, let us note that a natural next step consists in extending the ideas presented here to the general rotating Taub-NUT AdS black holes in d-spacetime dimension. It remains an interesting open question as to how the first law for d-dimensional black holes involves a variation of the Misner string angular momenta $J_s$ when multiple angular momenta are present. This task is left to future investigations.

\section*{Acknowledgements}

This work was supported by the NSF grant PHY-2012036 at Utah State University. The work of MJR is partially supported through the grants SEV-2016-0597 and PGC2018-095976-B-C21 from MCIU/AEI/FEDER, UE.


\appendix
\section{Killing–Yano tensor for Kerr NUT AdS}
\label{app:KY}

In this section we shall briefly describe the Killing–Yano tensor and their basic properties for Kerr Taub-NUT AdS black holes.

One of the most remarkable properties of the Kerr Taub-NUT AdS class of solutions, which is also inherited by its higher dimensional generalizations \cite{Chen:2006xh}, is the existence of hidden symmetries associated with the Killing–Yano tensor \cite{Kubiznak:2007kh}. Indeed, it is this tensor which is responsible for the ‘miraculous’ properties of this metric, including the integrability of geodesic motion or the separability of Hamilton–Jacobi and Klein– Gordon equations. As shown in \cite{Kubiznak:2007kh}, such a metric admits a hidden symmetry encoded in the CKY tensor  that takes the form
\bea
\textbf h&=&r\,dr \wedge\left[dt+\frac{d\phi}{\Xi}(2 n \cos\theta- a \sin^2\theta)\right] \nonumber \\
&& +\sin\theta \, (n+a \cos\theta) \,d\theta \wedge \left[ a\, \,dt-\frac{d\phi}{\Xi} \left(n^2+a^2+r^2 \right)\right]\,.
\eea
The Killing co-potential, relevant for the generalized Komar integrals defined in subsection \ref{sec:GSmarr} is defined via the CKY by $\omega= \textbf h/3$.

\end{document}